\documentclass[allclo,superscriptaddress,eqsecnum,amsfonts,showpacs]{revtex4}
\usepackage{epsfig}

\newcommand{\be}{\begin{equation}}
\newcommand{\ee}{\end{equation}}
\newcommand{\bea}{\begin{eqnarray}}
\newcommand{\eea}{\end{eqnarray}}

\newcommand{\ep}{i\varepsilon}
\newcommand{\nn}{\nonumber}

\begin{document}

\preprint{ \parbox{1.5in}{\leftline{hep-th/??????}}}

\title{On the  absence and presence of  pole(s) in fermion  propagator in 
the simple model with mass generation }

\author{V.~\v{S}auli}
\affiliation{CFTP and Departamento de F\'{\i}sica,
Instituto Superior T\'ecnico, Av. Rovisco Pais, 1049-001 Lisbon,
Portugal }
\author{F. Kleefeld}
\affiliation{Department of Theoretical Physics,
Nuclear Physics Institute, \v{R}e\v{z} near Prague, CZ-25068,
Czech Republic}

\begin{abstract}
Within the Pauli-Villars regularization technique 
the fermion propagator is studied in the framework of 
Schwinger-Dyson equations in the Euclidean space. 
Making the generalization of Fukuda and Kugo proposals,
the analytical continuation is performed into the timelike region
of fourmomenta. The massive gauge boson is considered and the fermion propagator pole 
structure is discussed in detail.
\end{abstract}

\maketitle
%

\section{Introduction}

The natural nonperturbative framework for the study of infrared properties of QCD Green's functions in the whole Minkowski space is the formalism of Schwinger-Dyson equations (SDEs). 
The complex of SDEs is an infinite tower of coupled integral equations which must be 
truncated to be tractable in practice. Most of the studies concerns the solution of SDEs in Quantum Chromodynamics \cite{CORNWALL} where it is believed they can give the correct description of chiral symmetry breaking and confinement. It is well known fact the the masses of the particles can be largely enhanced in  strong coupling theory like QCD. The property and difficulty to extract a correct picture of dynamical mass generation attracts much attention
in recent modeling of QCD \cite{CBRW2007,WIFIPE2007}, Technicolors  \cite{SHROCK}
and strong coupling models with spinless scalar bosons \cite{HOSEKBOYS}.
The models  in two later cases represent the dynamical alternatives to Standard model Higgs mechanism. These models can break symmetry but avoid the introduction of negative scalar boson mass term for this purpose.

In QCD, this is certainly   the asymptotic freedom and its associated regulating character of the ultraviolet modes  which makes chiral symmetry breaking physically meaningful. The same is assumed to be valid for Technicolor, however where the typical scale $\Lambda$ is readjusted about a few orders of magnitudes. In possible strong interacting Higgs extensions of Standard model the mechanism can be different since it could follow the ultraviolet completions of such models. In general, the mechanism is not completely understood but it  is expected it can be quite exotic as mother strong interacting theory can be  \cite{ARKANI}.  
Having no direct experimental hints which mechanism of mass generations the Nature has chosen, 
we can construct models based on effective interactions. The should describe correctly physics until some energy scales, above this, the underlying but unknown theory is assumed to be valid. 

In this paper we make a look on a simply approximated gauge theory of unspecified symmetry and regularize the high energy mode by Pauli-Villars regulators. The Pauli-Villars regulators then can be interpreted as the effect of physically hidden sector which effects is observed rather indirectly in the physical spectrum.  Very recently, the old fashionable idea 
of Wick and Lee \cite{WILE} has been reopened in \cite{GRCOWI2007}.
To this point we consider strongly interacting theory and  extend the technique of Fukuda-Kugo  solution \cite{FUKKUG} of gap equation to the case of Pauli-Villars regularization. We consider sufficiently heavy gauge boson with mass $\mu$ satisfying $\mu<\Lambda$, where $\Lambda$ is Pauli- Villars regulator. To see, what is the physical pole mass of fermion which received its mass dynamically, we make a continuation of the Schwinger-Dyson equation to the timelike axis of momenta $p^2>0$. We found rather nontrivial results depending on the mass ratio $\mu/\Lambda$.

The layout of the paper is as follows. Section II is devoted to the overview of Fukuda-Kugo 
 (timelike) ladder fermion SDE with hard cutoff regularization. The results with 
massless and massive gauge boson propagators inside the loop of SDE are obtained by the direct solution of the original integral equation.  
The absence of fermion poles has been actually  confirmed in the region predicted by the 
authors of \cite{FUKKUG}, however, in the light of our further discussion we  differ in the 
interpretation. In Section III we do the same for the equation with Pauli-Villars regulator.
 The numerical solution is shown in Section IV. Contrary to the naively regularized ladder  QED, the real fermion pole has been always  identified, at least for nonvanishing $\mu$. 
Note, the case  of  vanishing boson mass is more complicated issue and to this point the reader can see rather recent paper \cite{qcd} where the study of propagator of confined quarks has been attempted. Note for completeness and bearing in mind the fact that even gluons can reach their masses dynamically  \cite{HMOTAG}, the technical method proposed here can be straightforwardly used for "QCD with massive gluons". In Section V we discuss the possible form of 
singularities in the context of Lorentz invariance constraint. In Conclusion (Section VI) the 
basic results are summarized and further directions of research within this approach are 
outlined.

In this paper  we will use the following conventions:  the  positive variables $x,y$ will represent
 the square of momenta in the whole Minkowski space, i.e.  for instance $x=p^2$ for timelike 
momenta when $p^2>0$ while $x=-p^2$ for $p^2$ in spacelike region. Note, our metric is 
$g_{\mu\nu}=diag(1,-1,-1,-1)$. For purpose of  clarity we label the mass function $B$ as 
$B_s$ in the spacelike region of fourmomenta and as $B_t$ when evaluated for timelike
 fourmomentum (i.e. $B(p^2)=B_s(x)\theta(-p^2)+B_t(x)\theta(p^2)$).

\section{Maskawa-Nakajima/Fukuda-Kugo equation}

In this Section we review some basic facts on the Maskawa-Nakajima/Fukuda-Kugo equation. This equation  represents ladder approximation of the fermion SDE with the momentum integration is regulated by an upper boundary cutoff $\Lambda=p^2_{E,max}$.

 In parity conserving theory
 the fermion propagator $S$ can be characterized by two independent scalar function $A,B$ such that
 $S(p)=[A\not p-B]^{-1}$ (bare fermion propagator is $S_0=[\not p-m_0]^{-1}$).
In the ladder approximation, the equation for inverse of $S$ can be written
\be \label{zdar}
\not p A(p^2)-B(p^2)
=\not p -m_0-i g^2\int\frac{d^4q}{(2\pi)^4}\gamma_{\alpha}
G^{\alpha\beta}(p-q)S(q)\gamma_{\beta}
\ee
where $m_0$ is a bare mass and $ G^{\alpha\beta}$ is boson propagator. The equation Eq. (\ref{zdar}) and the classification of the solutions  has been discussed in  \cite{STREDOVEK} especially in Landau-like gauge for which the massive propagator reads
\be
G^{\alpha\beta}(q)=\frac{-g^{\alpha\beta}+\frac{q^{\alpha}q^{\beta}}{q^2}}{q^2-\mu^2+\ep}\, .
\ee
Few years later, Fukuda and Kugo  have found the solution for the timelike momenta for $\mu=0$. They observe no real pole for resulting fermion propagator. Hence the absence of free propagating mode has been interpreted as a sign for confinement at  that time.
First, disregarding the known deficiency of integral cutoff regularization scheme, we overview the method of solution and resolve (\ref{zdar}) also for nonzero $\mu$. Up to our knowledge such solution has  never been explicitly published in the literature. We leave the  question of  the reliability of 'would be' confining solution into the discussing section IV..

In the next we will  follow the paper \cite{FUKKUG}, using $A=1$ approximation and making Wick rotation and angular integration (see Appendix A)  we get 
\begin{eqnarray}\label{EU}
&& B_s(x)=m_0+\frac{3C}{4}
\int_0^{\Lambda^2}d y\, \frac{B_s(y)}{y+B_s^2(y)}
K(x,y,\mu^2)\,
\nn \\
&&K(x,y,\mu^2)=\frac{2y}{x+y+\mu^2+\sqrt{(x+y+\mu^2)^2-4x y}}
\end{eqnarray}
where $\Lambda$ is the cutoff and where $C=3g^2C_2(R)/4\pi^2$, $C_2(R)$ denotes the Casimir invariant of the quark representation ($C_2(R)=4/3$ for QCD). The equation (\ref{EU}) can be quite easily solved by the method of iteration avoiding thus inconvenient conversion to the nonlinear  differential equation.
When $m_0=0$ but  $B\ne 0$ we talk about dynamical chiral symmetry breaking.

Having kept the solution $B_s(y)$ for spacelike $y$
we can define the 'synthetic fermion mass' $\hat{B}(x)$ at timelike axis of fourmomenta  
such that $\hat{B}(x)=B_s(-x)$ is a solution of integral (\ref{EU}) for timelike x:
\begin{eqnarray}
&& \hat{B}(x)=m_0+\frac{3C}{4}
\int_0^{\Lambda^2}d y \, \frac{B_s(y)}{y+B_s^2(y)}
K(-x,y,\mu)\,
\nn \\
&&K(-x,y,\mu)=\frac{2y}{-x+y+\mu^2+\sqrt{-x+y+\mu^2)^2+4x y}}\, .
\end{eqnarray}
which represents correct continuation  until the first singularity
is crossed on the timelike axis of $p^2$.
The continuation   on the usual physical cut of  timelike momenta $x>(m+\mu)^2$ reads
\be
 B_t(x)=m_0+\hat{B}(x)-\frac{
3C}{4}
\int_0^{(\sqrt{x}-\mu)^2}d y \, \frac{B_t(y)}{y-B^2_t(y)+\ep}
X(x;y,\mu^2)\, \label{gen}
\ee
where $X(x;y,\mu^2)$ is the discontinuity of $K(-x,y,\mu^2)$ over the physical cut:
\be
X(x;y,\mu^2)=\frac{\sqrt{(-x-y+\mu^2)^2-4x y}}{x}\, ,
\ee
and note the presence of Feynman infinitesimal $\ep$, which has been omitted 
in the original paper \cite{FUKKUG}.

To see this continuation was correct let us assume the smoothness of $B$ at the vicinity of the pole $y=m^2$, i.e. $B_t^2(m^2)=m^2$. Then one can use the functional relation
\be
\frac{1}{{\cal{O}} \pm i\epsilon}=P.\, \frac{1}{\cal{O}}\mp i\pi\delta(\cal{O})
\ee
and obtain the dominant contribution to the absorptive part of mass function $B$
\be \label{ABS}
{\bf Im} B(x)=\frac{3C}{4}m\pi X(x;m^2,\mu^2)\Theta (x-(\mu+m)^2)\, ,
\ee
where $\Theta$ stands for the standard Heaviside step function. Clearly, (\ref{ABS})  represents  one loop perturbation theory result when  $C$ is a small parameter, $3C/4<<1$.

In general one have to solve the Eq. (\ref{gen}). Assuming smoothness of the function  $B$ the equation (\ref{gen}) can be rewritten as follows:
\bea \label{time}
B_t(x)&=&m_0+\hat{B}(x)- i\frac{3C}{4}\pi \sum_j \frac{m_j}{|1-2B_t B_t'|_j} X(x;m_j^2,\mu^2)\Theta (x-(\mu+m_j)^2)
-I(x)
\\
I(x)&=&\frac{3C}{4} P.
\int_0^{(\sqrt{x}-\mu)^2}d y \, \frac{B_t(y)}{y-B^2_t(y)}
X(x;y,\mu^2) \Theta(x-\mu^2)\,
\eea
where $j$ runs over the roots of Eq. $y-B^2_t(y)=0$ and where we have used the following abbreviation:
\be
|1-2B_t B_t'|_j=\left|1-2B_t(y)\frac{d B_t(y)}{d y}\right|_{y_j=m_j^2}=
\left|1-\frac{d B_t(y)}{d (y^{1/2})}\right|_{y_j=m_j^2}\, .
\ee

The Eq. (\ref{time}) represents integral inhomogeneous equations (even for zero $m_0$) with  the one singular kernel in the integrand of  $I$. It can be solved by  a standard numerical method and we provide some details in the Appendix B.  The dynamical chiral symmetry breaking is of great interest for us and  we will concern on these solutions in this paper. Inclusion of small explicit chiral breaking term is straightforward and we leave
this case aside of our interest.   

\begin{figure}
\centerline{\epsfig{figure=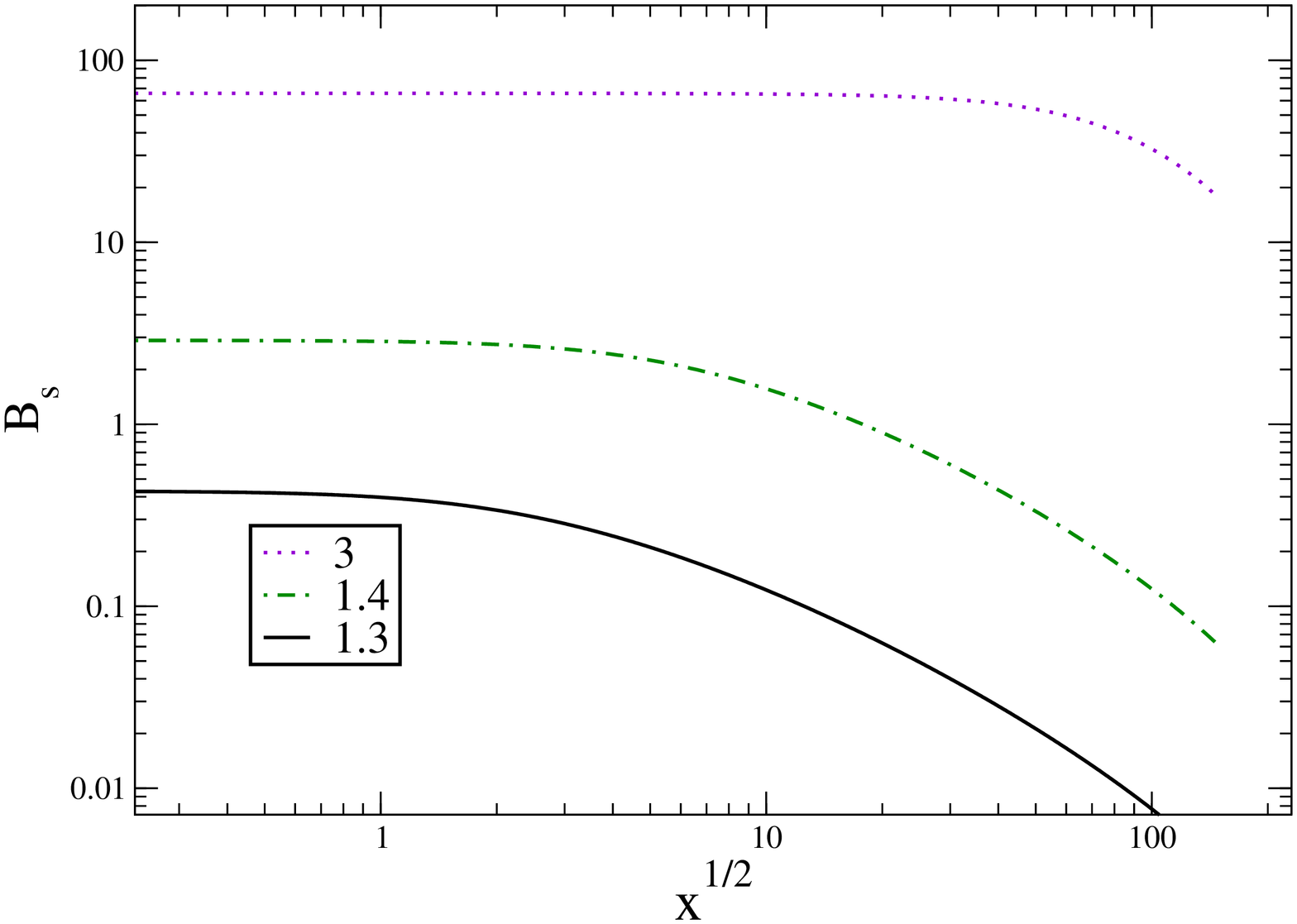,width=7.0truecm,height=7.0truecm,angle=270},
\epsfig{figure=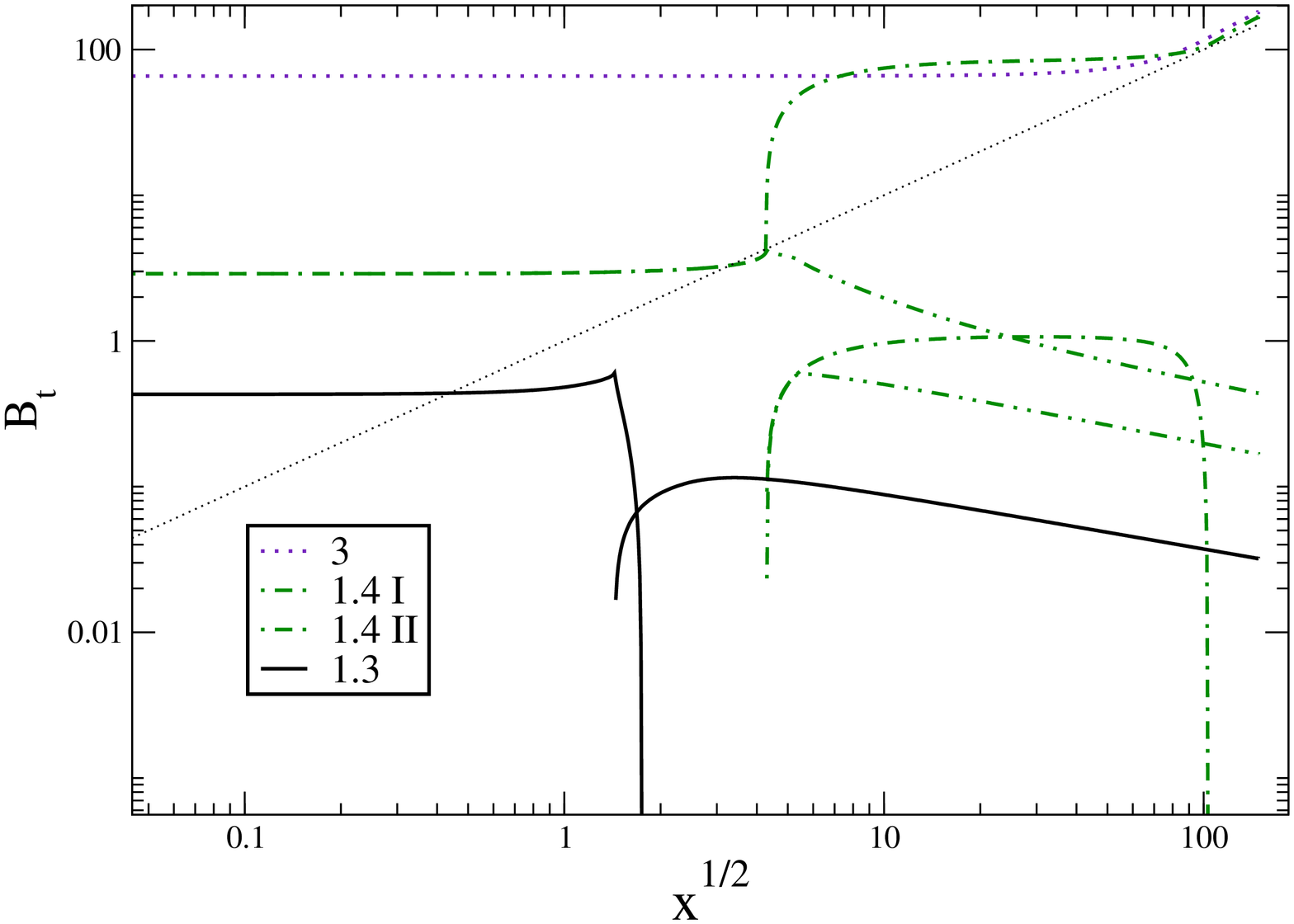,width=7.0truecm,height=7.0truecm,angle=270}} 
\caption[99]
{\label{fukuda} Dynamical mass function for spacelike (left) and timelike (right) region of fourmomenta for massive
 boson case and hard UV cutoff. The solutions are labeled by $\alpha=g^2/4\pi$. On the right panel the lines representing the 
imaginary part of $B$ becomes nonzero above the threshold. Notice, there are two solutions for the coupling $\alpha=1.4$. 
The pole mass are identified by the crossing of thin dotted line (the function $\sqrt(p^2)$).}
\end{figure}

 In order to scale the
 dimensionfull quantities  we take $\mu=1$ in arbitrary units. Further, the numerical hard cutoff 
is adjusted to be $\Lambda=e^5\mu\simeq 148\mu$ . Its value  constraints  
the  meaningful region of the square of fourmomenta as $p^2\in (-\Lambda^2,\Lambda^2)$. The numerical solutions for various coupling constant are presented in Fig. 1.

 As  expected already from the discussion 
in \cite{FUKKUG} there are several phases of the theory. 
They have been confirmed and actually found also here. They are as follows: 

{\bf I.} Chiral symmetric phase- for the coupling bellow the critical value $\alpha_c$,
 there is only trivial solution $B=0$. 

{\bf II.} Chiral symmetry breaking phase, where the propagator has a real pole.
 In this case we get also nontrivial solution for  $B$. For our parameters $\mu,\Lambda$
 this phase is characterized by the coupling strength $\alpha\in (\alpha_c,1.4)$  and the pole mass
 $m$ is at most of order $\mu$. In fact, we can observe and have found two solution for 
$\alpha=1.4$. Although numerically  they both cut the axis $\sqrt{p^2}$, the one of them only sweep the 
region under the diagonal and their timelike high energy asymptotic crucially differ. 
Of course, the question of uniqueness of such continuation naturally arises.

{\bf III.} Chiral symmetry breaking phase above $\alpha=1.4$, where the mass ratio  $m/\mu\simeq 4.2 $
 (and $B(0)\simeq 3\mu$) is achieved, the pole and associate analytical cut vanish and the mass function
 blows to 'infinity' like $\sqrt(p^2)$ for large $p^2$.
Note, no sign of such transition is  observed in the spacelike regime. 
Clearly the spacelike solution does not reflect  dramatic changes that happen in timelike
part of the Minkowski space.  
The  solutions in the regime III. have been originally  called 'confining' because of the absence of 
real pole and associate particle production threshold.  In fact we have to be more careful 
with the interpretation, particularly because the Wick rotation is not allowed, 
 the rotation of the contour cross the singularity at timelike momentum infinity. 
Clearly, in this case the hard cutoff regularization of the momentum integral does not
 commute with the Wick rotation  and  there is neither confidence that we obtain a true solution
 in the whole Minkowski space.  This is  because we are dealing with non-asymptotic
 free theory where 
the result are not independent on the UV integral cutoff. We argue here,
 the "would be" confining solution would be meaningful only when the
 theory is defined in the Euclidean space from the beginning. Clearly, this is not the case of ladder QED and the observed nontrivial solution can be  an artifact of given scheme. Although we suppose that these things are rather well known, we 
 discussed them  explicitly here for the purpose of clarity.

\subsection{Massless vector boson}

The solution for massless vector boson has been firstly obtain in \cite{FUKKUG}. 
For this purpose the authors of the paper  \cite{FUKKUG}  transform the integral equation into the differential one which has
 been solved numerically. Here we confirm their solution by  direct solution of the original integral
 equation. Taking a limit $\mu\rightarrow 0$ in Eq's (\ref{EU}) and (\ref{time}) is straightforward.
The SDE for selfenergy in the massless case $(m_0=\mu=0)$ then reads 
\be \label{zerone}
 B_s(x)=\int_0^x d y \left(\frac{y}{x}-1\right) \frac{B_s(y)}{y+B^2_s(y)}
+\frac{<\psi\bar{\psi}>_{\Lambda}}{N_c}
\ee
\be \label{zertwo}
 B_t(x)=\frac{<\psi\bar{\psi}>_{\Lambda}}{N_c}-\frac{3C}{4}
\int_0^x d y \left(1-\frac{y}{x}\right) \frac{B_t(y)}{y-B^2_t(y)}
\ee
where $N_c$ is the number of colors and where we have already omitted intensifimal imaginary prefactor 
$\ep$ since as it is known that the equation  $y-B^2_t(y)=0$ has no real roots in hard cutoff theory.
The inhomogeneous term in (\ref{zertwo}) is simple constant-the usual fermion condensate: 
\be 
\hat{B}=\frac{<\psi\bar{\psi}>_{\Lambda}}{N_c}=\frac{3C}{4}\int_0^{\Lambda^2}\frac{B_s(y)}{y+B^2_s(y)}
\ee 

\begin{figure}
\parbox{8cm}{\epsfig{figure=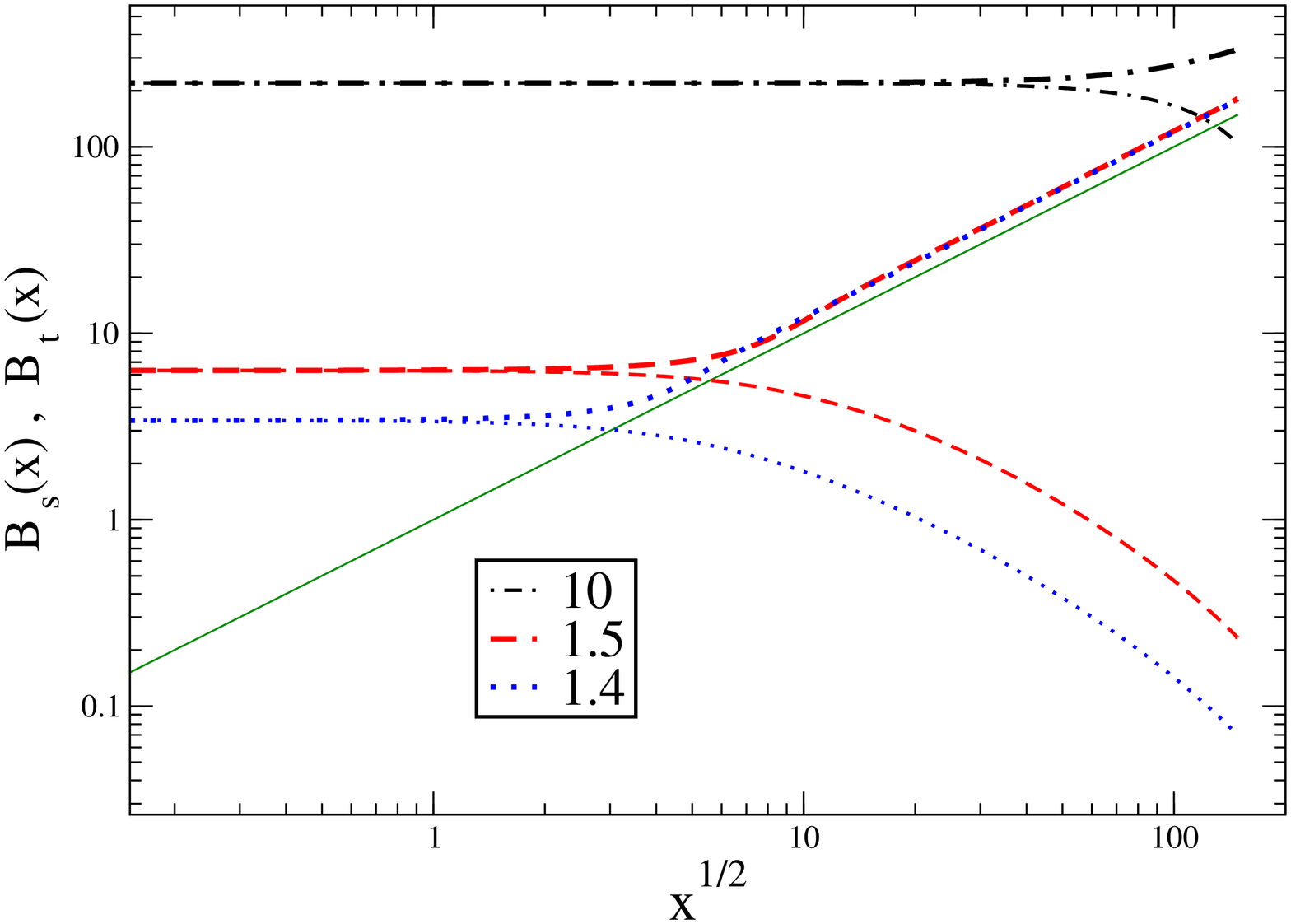,width=8truecm,height=8truecm,angle=270}
\caption[90]{Fermion dynamical mass function in ladder approximation with hard cutoff $\Lambda^2=e^{10}$ (in arbitrary units) for the three value of the coupling constant
$\alpha=1.4,1.5,10$. The thin dotted line represents $\sqrt{p^2}$, the solution for timelike momenta are labeled by thick 'upper' lines, the spacelike solutions are added for the comparison (decreasing lines of the same type belongs to the same coupling, clearly $B_s $ and $B_t$ are identical  for $x=p^2=0$)\label{massless} } }
\parbox{8cm}{\epsfig{figure=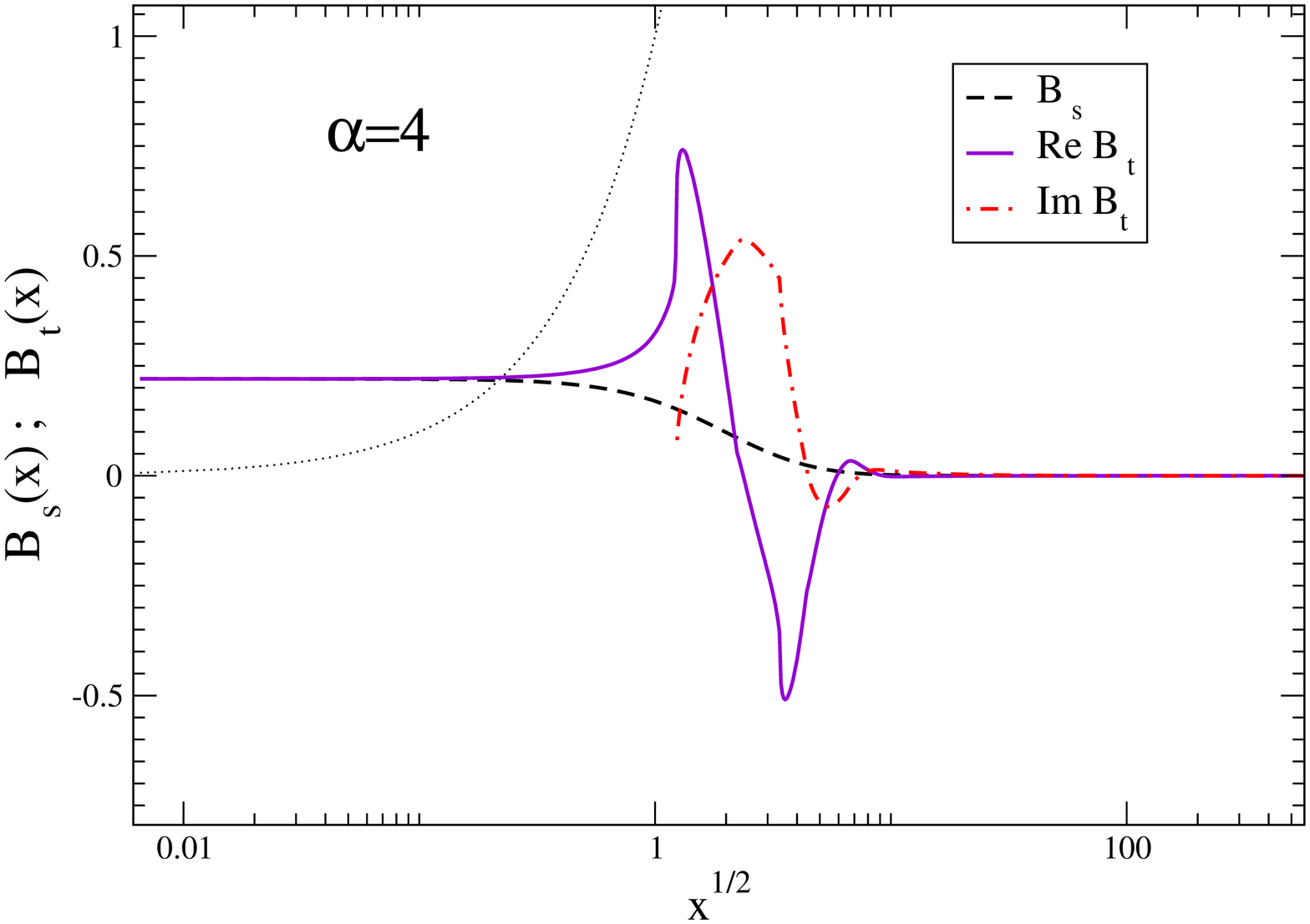,width=8truecm,height=8truecm,angle=270}
\caption[caption]{Dynamical quark mass for  $\alpha=4$ as described in the text.
\label{pauli4}} }
\end{figure}

The numerical results are shown in Fig. \ref{massless} for three distinct values of the coupling constant.
Similarly to previously discussed region III of massive boson case the results could be interpreted with a great care. Again the Wick rotation invalidates because of naive regularization scheme and the resulting timelike high momentum behaviour can be  an artifact  of  inappropriate calculation scheme.

\section{Dynamical mass generation with Pauli-Villars regularization}

In this Section we consider the fermion gap equation regularized by Pauli-Villars regulator function. Such construction is very straightforward in our approximation since it achieved by 
the replacement of the gauge boson propagator by the sum of ordinary , i.e. original one propagator, and the "wrong sign" boson propagator but with Pauli-Villars mass $\Lambda$ instead 
of $\mu$. Explicitly, we make the following in the kernel of the gap equation

\be  \label{aprc}
\frac{-g^{\mu\nu}+\frac{k^{\mu}k^{\nu}}{k^2}}{k^2-\mu^2+\ep}
\rightarrow \left[-g^{\mu\nu}+\frac{k^{\mu}k^{\nu}}{k^2}\right]
\left[\frac{1}{k^2-\mu^2+\ep}-\frac{1}{k^2-\Lambda^2+\ep}\right] \, .
\ee
Within a finite $\Lambda$ we do not consider the hard cutoff, which is formally  removed from now. 

and where we have also introduced effective gluon mass $\mu$.

In order  to safely define the  pole fermion mass we  consider massive boson
 case in presented study. We suppose that the timelike singularity can be drastically changed due to  the presence of strong interaction and we leave this complicated problem for separate numerical search and discussion elsewhere.
However, as usually we assume no dynamical singularities in the fermion  propagator in the first and the third quadrant
of complex $p^2$ plane. It not even allows to
perform standard Wick rotation, but the main advantage of presented method is that  we can readily follow the Fukuda-Kugo trick and make the continuation to the timelike axis. 

Doing this explicitly we get  for spacelike solution:
\begin{eqnarray}
&& B_s(x)=m_0+\frac{\alpha}{\pi}
\int_0^{\infty}d y \, \frac{B_s(y)}{y+B_s^2(y)}
K(x,y)\,
\nn \\
&&K(x,y)= \left[ K(x,y,\mu^2)-K(x,y,\Lambda^2)\right]
\nn \\
&&=\frac{1}{2x}\{\mu^2-\Lambda^2-\sqrt{(x+y+\mu^2)^2-4x y}+\sqrt{(x+y+\Lambda^2)^2-4x y}\}
\end{eqnarray}
and the continuation  on the analyticity cut reads
\bea  \label{pauli}
&&B_t(x)=\hat{B}(x)-I(x)+ i\alpha \sum_j \frac{X(x;m_j^2,\mu^2)\Theta (x-(\mu+m_j)^2)-
X(x;m_j^2,\Lambda^2)\Theta (x-(\Lambda+m_j)^2)}{\left|1-\frac{d B_t(y)}{d (y^{1/2})}\right|_{y_j=m_j^2}}\, ,
 \\
&& \hat{B}(x)=m_{0}+\frac{\alpha}{\pi}
\int_0^{\infty}d y \, \frac{B_s(y)}{y+B_s^2(y)}
K(-x,y)\, ,\nn \\
&&I(x)=\frac{\alpha}{\pi} P.
\int_0^{(\sqrt{x}-\mu)^2}d y \, \frac{B_t(y)}{y-B^2_t(y)}
X(x;y,\mu^2) \Theta(x-\mu^2)
-\frac{\alpha}{\pi} P.
\int_0^{(\sqrt{x}-\Lambda)^2}d y \, \frac{B_t(y)}{y-B^2_t(y)}
X(x;y,\Lambda^2) \Theta(x-\Lambda^2)\, ,
\nn \\
&&K(-x,y)=K(-x,y,\mu^2)-K(-x,y,\Lambda^2)
\nn \\
&&=\frac{1}{-2x}\{\mu^2-\Lambda^2-\sqrt{(-x+y+\mu^2)^2+4x y}+\sqrt{(-x+y+\Lambda^2)^2+4x y}\} \, .
\nn \\
\eea
The equations (\ref{pauli}) have been solved numerically by the method of iterations, 
some  useful details concerning the numerical treatment can be found in the Appendix.

Remind here  known feature of  QCD scaling: the constituent quark masses are of the same size as the QCD scale, i.e.  $B(0)\simeq \Lambda$. In this paper, for any reason, we freely focus on a  larger region of parameter space providing us the results for softer coupling $B(0)< \Lambda$ and stronger $B(0) > \Lambda$ couplings as well as. We plot sample of   numerical  solutions in Fig. 3-6.
For most of the solutions we fix the mass ratio to be $\Lambda^2/\mu^2=10$, the exception is explicitly mentioned. The critical value of the coupling has been identified $\alpha_c\simeq3.8$, bellow that we do observe the trivial solution only. In Fig. 3 the numerical solution is plotted for the coupling strength $\alpha=4.0$. Being rather close to the critical value $\alpha_c$, the quark propagator has a one real pole at some  point $m$ which value is very closed to the infrared mass $B(0)$. Without any doubt,  the constituent mass can be freely identified with the infrared mass or with  the pole one. In Fig 4. the solution for $\alpha=5$ is shown and we observe that propagator develops two real poles under the first branch point. The absorptive part of $B$ is largely enhanced because the both differentiations  $dB(m)/d m$ are not so far from 1 (see Eq.   
(\ref{pauli})). Such solution is possible since the boson is massive, however it  may be an artifact of our approximation. The observation of two pole solutions is in agreement with
similar observation mad in \cite{SAULI}, however we regard this as a curiosity rather then possible physical scenario. Increasing the coupling further, then the second  pole  vanishes at some point and we retain with the one real pole solution again. This situation is exhibited in Fig 5. . Note, the propagator becomes largely enhance bellow $m$, since the function $B_t$ lies very closed to the diagonal of $\sqrt{p^2}, B_t$ graph. Contrary to the case of solutions with two poles, the imaginary part of $B$ appears to be suppressed now, because of differentiation $B'(m)<0$. In that case the pole mass largely differ from the infrared value $B(0)$, it is typically few times higher.

\begin{figure}
\parbox{8cm}{\vspace{-1.0cm}
\centerline{\epsfig{figure=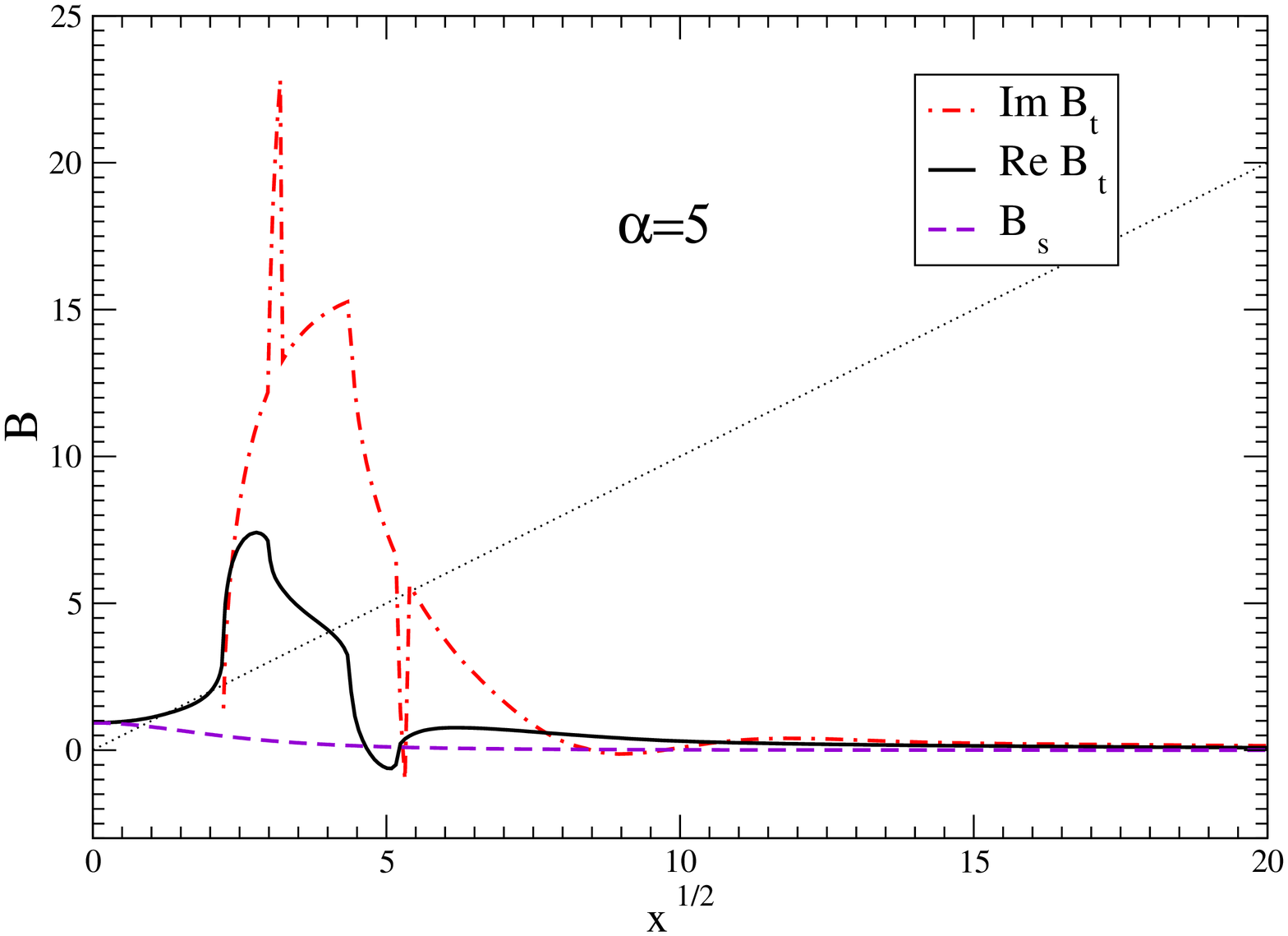,width=8truecm,height=8truecm,angle=270}}

\caption[caption]{Dynamical  fermion mass for $\alpha=5$ \\
 and $\mu/\Lambda=\sqrt{0.1}$.} \label{pauli5}
}
\parbox{8cm}{\centerline{\vspace{3mm}\epsfig{figure=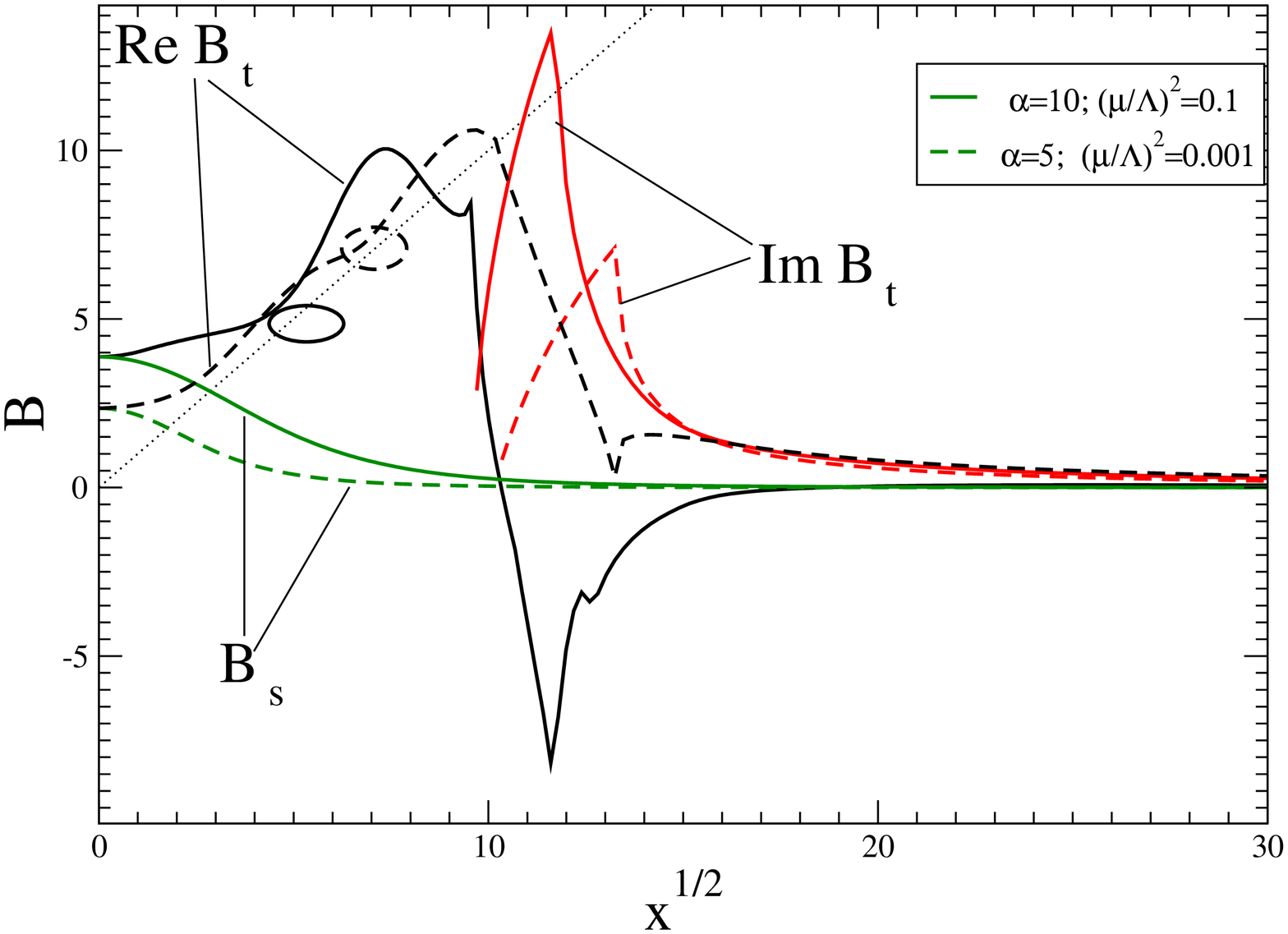,width=8truecm,height=8truecm,angle=270}}
\caption[caption]{\hspace{2mm}Dynamical  masses  for 'super-strong' coupling as described in the text.  Solid lines represent  result for $\alpha=10$ and $\mu/\Lambda=\sqrt{0.1}$,  dashed line stands  for $\alpha=5$ and $\mu/\Lambda=\sqrt{0.001}$   } \label{pauli10}}
\end{figure}
\vspace{1cm}

\section{Discussion of the results}

In this section we continue the discussion of the results and qualitatively compare with the already known results presented in  literature. We also discussed the  location of the poles in the complex plane and its relations with the violation of Lorentz invariance.
\begin{figure}
\centerline{\hspace{-8.cm}
\epsfig{figure=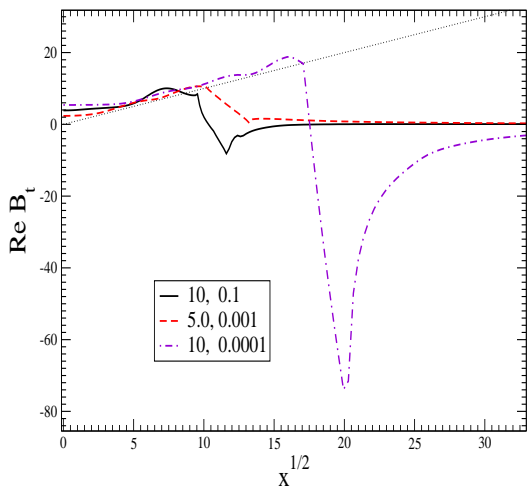,width=8truecm,height=8truecm,angle=270}}
\caption[caption]{
Real part of the dynamical mass function  \\
for $x=p^2$. The lines  represent the models  \\
characterized by the couple of  parameters $\alpha$ and the ratio 
$\mu^2/\Lambda^2$ as shown in the box.  } \label{time10}
\end{figure}

In QCD with chiral quarks the asymptotic freedom  ensures the finiteness  of dynamically generated masses without any use of  regularization scheme. In our case the infinity is regulated by the Pauli-Villars term induced by auxiliary unphysical field added to the Lagrangian, hence the mass function would be finite even when the bare mass is added. As we have  checked explicitly for small bare mass,  it  affect the ultraviolet behaviour of the mass function but has a little effect in the infrared and the singularity structure remains the same.    
The main result of our paper is the numerical observation of the fact that the  real singularity is never absent in our translation invariance preserving calculation scheme.  
The singularity is a real pole for large parameter space of $\alpha, \mu$ and $\Lambda$
However approaching the limit of vanishing boson mass $\mu$, the pole is going to meet the branch point and the situation is getting more complicated. In  Fig. 5 the result is plotted for  a small ratio $\mu^2/\lambda^2=0.001$. We should note that the exact massless limit may not be well defined
in approach presented here so that Fukuda-Kugo continuation method is not fully justified in this case. Recall, the functional used in our equations is defined on the set of differentiable functions, while in contradiction,  the differentiation $B'(m)$ may not exist in the exact massless case $\mu=0$. The zero boson mass case will be studied elsewhere due to its own
serious physical consequences. 

Stating without proof, we suppose that true mass generation and  the evidence of nontrivial pole of the fermion propagator go hand by hand with the property of asymptotic freedom (decreasing of effective coupling at ultraviolet), whilst the absence of real pole as happened to  the propagator function in Fukuda-Kugo equation is artifact of improper calculation scheme (hard cutoff regularization). Further note, the numerical results obtained here qualitatively correspond with the results of the papers   \cite{SAULI}, where the same model was studied in the Minkowski space for the first time. 
At this place we discuss some common and some distinct features of solutions presented here and in the paper \cite{SAULI}. In both  approaches the validity of Wick rotation is assumed. In other words, no (dynamically generated) complex  singularities are assumed in the first and the third quadrant of $p^2$ complex plane. The main distinction is the analyticity assumption  explicitly used in the paper \cite{SAULI}, where it was assumed that the propagator is holomorphic in the whole  $p^2$ complex plane up to a real positive semi-axis and that the propagator satisfy certain integral (generalized spectral) representation. To this point, there is no similar assumption in the approach developed by Fukuda-Kugo and followed in presented paper.
In the method presented here the correlator at timelike regime of fourmomenta
 is built on a base of the knowledge of  its spacelike counter-partner.
However, being possibly out of the domain of analyticity we should be aware that such continuation is by no mean unique.
On the other hand , there is also reasonable quantitative agreement with \cite{SAULI}, especially when we are not far form the critical coupling characterizing dynamical chiral symmetry breaking. Actually, depending on the coupling strength $\alpha$ the propagator develops one or two real poles with corresponding branch points providing their values are in an approximate agreement with  \cite{SAULI}.

In the literature there is a certain effort to guess the complex 
structure of QCD Green's function by the method of continuation of Euclidean result or by indirect reading from the  phenomenological consequences which would follow form  particularly assumed singularity structure of Greens functions.
Complex conjugated singularities of quark  and fermion propagator in planar 
strong coupling QED  have been considered in the context of confinement 
\cite{FAKTNE} and in PT symmetric Quantum Field theory \cite{BSBCW2004}.
Additionally recent studies have modeled Euclidean space lattice data with propagators that have complex conjugate singularities \cite{BPRT2003,ALDEFIMA2003}.
More phenomenologically, the meson bound states \cite{BPT2003} and the parton distributions
\cite{TDM2003} has been calculated with a quark consisting of pairs of complex four momenta.
In many, if not at all of these recent studies, the fact that complex singularity structure can affect the  Lorentz invariance of the theory has been overlooked. Therefore, in what follow we mention the question of Wick rotation, location of complex poles and possible lost of Lorentz invariance  in order to pay  attention for.

Actually, increasing the  coupling strength one can expect that  apart of the real pole  a new complex singularities appear.
The area of real part of $p^2$  where we can expect new complex singularities in the propagator function is indicated by the ellipses in  Fig. \ref{time10}. In recent, we are not able to estimate the  characteristic of the behaviour and/or  the position of these complex singularities with good confidence. Instead of this, we would like to present the argument which largely enforces our believe in the validity of Wick rotation in a case of full realistic solution of Schwinger-Dyson equations. Recall at this place, there is known relation between location of complex singularities of Greens functions and the Lorentz invariance of the theory. Indeed, its more than 30 years known that complex singularities  located simultaneously on both side of the real axis of $p^2$ automatically generates Lorentz violating peaces in the propagator itself \cite{SUDARSAN}.
In this case the naive Wick rotation invalidates and one has to account the complex singularities in a suitable manner. Actually,  assuming the complex poles (with non vanishing Im part) one can  exhibit the existence of  Lorentz violating pieces originally not expected in. The evaluation of the appropriate Feynman integral has been exhibited for the case of complex conjugated poles
 in \cite{SUDARSAN} (some prefactor has been corrected in the paper \cite{NEWNAKAN}, the actual evaluation has been performed for scalar field content only, the extension to the  loop integral with internal fermion links is rather straightforward) due to the rather different reasons. The generalization to more general locations of singularities were already  discussed in \cite{SUDARSAN}. From these arguments it follows: if the true exact solution respect the Lorentz invariance of the theory then  selfconsistence of the Schwinger-Dyson equation solution for quark propagator can not involve complex poles on the both sides of the real axis of the square of the fourmomentum. In Lorentz invariant theory, the Wick rotation is indirectly justified in this way.


\section{Conclusion}

The model we have described , is of course not true QCD and hence we do not claim that the fermion propagator we have studied here is truly representative of the singularity structure of the quark propagator. In a more realistic models, which include the dressed vertex unction and/or higher order skeleton graphs, one expects that the details of the propagator structure would be different.
It should also incorporate the aspect of asymptotic freedom in a more proper way. However, our study demonstrates that singularity structure of the quark propagator
is very likely dominated by the real singularity and we expect a quantitative but not a qualitative changes when the approximations improve.
We discuss possible singularity structure in the entire complex plane of momenta. 
We argue, in the Lorentz covariant formulation of QCD (and any QFT) the singularities of the propagators (and of integral kernel of gap equations at all) must appear only on the one side -upper or down- separately for the left and the right half complex plane. Otherwise we necessarily sacrify Lorentz invariance of the theory. This argument strongly support the validity of Wick rotations in Lorentz invariant theory.

\section{Acknowledgments}

V.~\v S and  was supported 
by the grant GA CR 202/06/0746 
F.K. was supported by  M\v{S}MT, ``Doppler Institute" project Nr. LC06002. 

\appendix

\section{ Angular integration}

The following integral: 
\be
\int_{-1}^{1} dz \frac{\sqrt{1-z^2}}{z-a}=
\pi(-a+\sqrt{a^2-1})\, ,
\ee
is used in order to perform the angular integration in SDEs.

\section{Principle value integration}

\subsection{$I$ for Fukuda-Kugo equation }

  SDE for timelike momentum (\ref{time}) represents the complex inhomogeneous integral  equation with singular kernel.  In this appendix we describe some details how to numerically deal with.
The integral to be evaluated reads
\be \label{kockud}
I(x)=\frac{3C}{4} P.
\int_0^{(\sqrt{x}-\mu)^2}d y \, \frac{B_t(y)}{y-B^2_t(y)}
X(x;y,\mu^2) \Theta(x-\mu^2)\,
\ee
noting that the other terms in the SDE represent regular integrals with a smooth kernel.

Let us assume that we have made a good guess of the value of the pole mass $m$,  then it is convenient to write down the expression (\ref{kockud}) separately for various regime of momenta
term $I$ equivalently as follows
\be 
I(x)=0 \, \hspace{2cm} \, x<\mu^2
\ee

The function $B(y)$ is real bellow the thresholds and the kernel is regular  for $x$ bellow the threshold, hence  there is no need to denote $P.$ in front of integral since
$y<m^2$ for $\sqrt(x)<m+\mu$. At the point $\sqrt(x)=m+\mu$ the kernel singularity is suppressed by vanishing function $X$, thus we can safely write
\be
I(x)=\frac{3C}{4} 
\int_0^{(\sqrt{x}-\mu)^2}d y \, \frac{B_t(y)}{y-B^2_t(y)}
X(x;y,\mu^2)\, \,\hspace{2cm} \mu^2<x<(\mu+m)^2
\ee
For a larger $x$ we necessarily cross the singularity in the kernel, however $B(y)$ remains real to the threshold and we consider this regime separately:
\be \label{B4}
I(x)=\frac{3C}{4} 
P. \int_0^{(\sqrt{x}-\mu)^2}d y \, \frac{B_t(y)}{y-B^2_t(y)}
X(x;y,\mu^2)\, , \,\hspace{2cm} (\mu+m)^2<x<(2\mu+m)^2
\ee
Increasing $x$ the kernel become complex and we divide the integral to the  P. value integration over the real $B$ and  to the regular integration over the regular kernel with  complex $B$:
\bea
I(x)&=&\frac{3C}{4} 
P. \int_0^{(m+\mu)^2}d y \, \frac{B_t(y)}{y-B^2_t(y)}
X(x;y,\mu^2)\, 
\nn \\
&+&
\frac{3C}{4} 
 \int_{(m+\mu)^2}^{(\sqrt{x}+\mu)^2}d y X(x;y,\mu^2)  
\left[\Sigma_R
\frac{y-\Sigma_R^2-\Sigma_I^2}
{\left(y-\Sigma_R^2+\Sigma_I^2\right)^2+4\Sigma_R^2\Sigma_I^2}
+i\, \Sigma_I
\frac{y+\Sigma_R^2+\Sigma_I^2}
{\left(y-\Sigma_R^2+\Sigma_I^2\right)^2+4\Sigma_R^2\Sigma_I^2}\right]\, ,
\nn \\
&&{\mbox{for}} \hspace{1cm}\, x>(2\mu+m)^2\,  ,
\eea
where we have used following shorthand notation
\be
\Sigma_R={\bf Re} B_t(y); \, \, \Sigma_I={\bf Im} B_t(y).
\ee

To avoid some unwanted numerical fluctuations which usually stem from asymmetric distribution of mesh points when P. integration is numerically performed, we use a standard trick. Consider for this purpose the third of considered momentum regime where x runs over the 
interval $((\mu+m)^2,(2\mu+m)^2)$ (see (\ref{B4})) . In our numerical treatment the integral is replaced by a discrete sum with the appropriate (Gaussian) weights, i.e. 
\be
I(x)=\frac{3C}{4} 
P. \int_0^{(\sqrt{x}-\mu)^2}d y \, \frac{B_t(y)}{y-B^2_t(y)}
X(x;y,\mu^2)\,\rightarrow \frac{3C}{4}\sum_j 
 w(y_j) \, \frac{B_t(y_j)}{y_j-B^2_t(y_j)}
X(x;y_j,\mu^2)\, .
\ee

The numerical fluctuations are subtracted by the following trick 
\bea
&& \frac{3C}{4}\sum_j 
 w(y_j) \, \frac{B_t(y_j)}{y_j-B^2_t(y_j)}
X(x;y_j,\mu^2)
\nn \\
&&=\frac{3C}{4}\left\{\sum_j w(y_j)\left[
  \, \frac{B_t(y_j)}{y_j-B^2_t(y_j)}
X(x;y_j,\mu^2)- 
  \frac{m X(x;m^2,\mu)}{y_j-m^2}\right]
+ X(x;m^2,\mu^2)\log{|\frac{(\sqrt{x}-\mu)^2-m^2}{-m^2}|}\right\}\, ,
\eea
where the later two terms vanishes in the exact continuum limit.  
We have found this approach is very stable and the pole mass $m$ can be identified after 
few runs of the iteration program. In order to achieve better stability, the actual search of the root of equation $B(x)-x=0$ has been performed by hand for each solution.  After making a few iterations in $m_j$  then the achieved numerical accuracy  of our  search is estimated by   $\Delta m\simeq step/m$ where 'step' means the difference of two neighboring points at vicinity of $m$.
Likewise in the case of SDE in Euclidean space,  the integrals were cut by the ultraviolet cutoff.

\subsection{Kernel with Pauli-Villars propagator added}

Generalizing to the case of Eq. (\ref{pauli}) is rather straightforward. The principal value integration is treated in the same fashion as in the case of   Fukuda-Kugo equation.
Now the relevant term reads
\be
I(x)=I_{\mu}(x)-I_{\Lambda}(x) \, ,
\ee
where $I_{\mu}$ and $I_{\Lambda}$  are the integrals considered previously, but where $\mu$ is replaced by QCD scale $\Lambda_{QCD}$ in the later case. Contrary to the previous case the integral is insensitive to the value of upper boundary $\Lambda_H$, when it $\Lambda_H>\Lambda_{QCD}$

For completeness we list the function   $I_{\Lambda}$ bellow

\bea 
I_{\Lambda}(x)&=&0 , \hspace{3cm} x<{\Lambda}^2
\nn \\
I_{\Lambda}(x)&=&\frac{3C}{4} 
\int_0^{(\sqrt{x}-\Lambda)^2}d y \, \frac{B_t(y)}{y-B^2_t(y)}
X(x;y,\Lambda^2)\, ,\hspace{2cm} \Lambda^2<x<(\Lambda+m)^2
\nn \\ 
%
I_{\Lambda}(x)&=&\frac{3C}{4} 
P. \int_0^{(\sqrt{x}-\Lambda)^2}d y \, \frac{B_t(y)}{y-B^2_t(y)}
X(x;y,\Lambda^2)\, ,\hspace{2cm} (\Lambda+m)^2<x<(\mu+\Lambda+m)^2
\nn \\
I_{\Lambda}(x)&=&\frac{3C}{4} 
P. \int_0^{(m+\mu)^2}d y \, \frac{B_t(y)}{y-B^2_t(y)}
X(x;y,\mu^2)\,
\nn \\ 
&+&
\frac{3C}{4} 
 \int_{(m+\mu)^2}^{(\sqrt{x}-\Lambda)^2}d y X(x;y,\Lambda^2)  
\left[\Sigma_R
\frac{y-\Sigma_R^2-\Sigma_I^2}
{\left(y-\Sigma_R^2+\Sigma_I^2\right)^2+4\Sigma_R^2\Sigma_I^2}
+i\, \Sigma_I
\frac{y+\Sigma_R^2+\Sigma_I^2}
{\left(y-\Sigma_R^2+\Sigma_I^2\right)^2+4\Sigma_R^2\Sigma_I^2}\right]\, ,
\nn \\
&&\mbox{for} \hspace{2cm} \, x>(\mu+\Lambda+m)^2\, 
\eea
 The treatment of principal value integrals
is the same as in the previous case of Fukuda-Kugo equation.



\begin{thebibliography}{99}

%
\bibitem{CORNWALL}
Mandelstam S., Phys. Rev. {\bf D20}, 3223 (1979);
Higashijima K., Phys. Rev. {\bf D29}, 1228 (1984);
Brown N.,  Pennington M. R., Phys. Rev. {\bf D39}; 2723 (1989);
L. von Smekal, A. hauck and R. Alkofer, Ann. Phys. {\bf 267}; 1 (1998);
Atkinson D., Bloch J.C.R., Phys. Rev. {\bf D58}, 094036n (1998);
Cornwall J.M., Phys. Rev. {\bf D26}, 1453 (1982);
C. S. Fischer, J. Phys.{\bf G32}, R253 (2006).
%
\bibitem{CBRW2007}
L. Chang, Y.-X. Liu, M.S. Bhagwat, C.D. Roberts and S.V. Wright, Phys. Rev. {\bf C75}, 015201 (2007), [nucl-th/0605058].
%
\bibitem{WIFIPE2007}
R. Williams, C. S. Fischer, M.R. Pennnigton, arXiv:0704.2296, hep-ph. 
%
\bibitem{SHROCK}
M. Kurachi, R. Shrock, JHEP 0612, 034 (2006). 
%
\bibitem{HOSEKBOYS}
P. Benes, T. Brauner, J. Hosek, Phys. Rev. {\bf D75}, 056003 (2007).
%
\bibitem{ARKANI}
N. Arkani-Hamed, A.G. Cohen, H. Georgi, Phys. Lett. {\bf B513}, 232 (2001);
N. Arkani-Hamed, A.G. Cohen, T. Gregoire, E. Katz, A.E. Nelson, J.G. Wacker,
JHEP {\bf 0208},021 (2002);
B. Grinstein, M. Trott, arXiv:0704.1505.
%
\bibitem{WILE}
T.D. Lee and G.C. Wick, Nucl. Phys. B{\bf 9}, 209 (1969);
T.D. Lee and G.C. Wick, Nucl. Phys. D{\bf 2}, 1033 (1970).
%
\bibitem{GRCOWI2007}
B. Grinstein, D. O`Connell, and M.B. Wise,
arXiv:0704.1845 8hep-ph]
%
%
\bibitem{FUKKUG}
Fukuda R., Kugo T. , Nucl. Phys. {B 117}, 250 (1976).
%
\bibitem{qcd}
V. Sauli, arXiv:0704.2566 .

%
%
\bibitem{STREDOVEK}
K. Johnson, M. baker and R. Willey, Phys. Rev. {\bf 136} B1111 (1964);
K. Johnson and M. Baker,  Phys. Rev.{\bf D8}, 1110 (1974);
T. Maskawa and H.Nakajima, prog. Theor. Phys. {\bf 52}, 1326 (1974); {\bf 54}, 860 (1975).
%
\bibitem{HMOTAG}
%
M.~J.~Lavelle and M.~Schaden,
Phys.\ Lett.\ B \textbf{208}, 297 (1988);
%
F.~V.~Gubarev, L.~Stodolsky and V.~I.~Zakharov,
Phys.\ Rev.\ Lett.\ \textbf{86}, 2220 (2001);
%
H.~Verschelde, K.~Knecht, K.~Van Acoleyen and
M.~Vanderkelen,
Phys.\ Lett.\ B \textbf{516}, 307 (2001);
%
K.~I.~Kondo, T.~Murakami, T.~Shinohara and T.~Imai,
Phys.\ Rev.\ D \textbf{65}, 085034 (2002);
%
U.~Ellwanger and N.~Wschebor,
Int.\ J.\ Mod.\ Phys.\ A \textbf{18}, 1595 (2003);
%
 D.~Dudal, H.~Verschelde, J.~A.~Gracey,
V.~E.~R.~Lemes, M.~S.~Sarandy, R.~F.~Sobreiro and S.~P.~Sorella,
JHEP \textbf{0401}, 044 (2004);
%
A.~C.~Aguilar, A.~A.~Natale and P.~S.~Rodrigues da Silva,
Phys.\ Rev.\ Lett.\ \textbf{90}, 152001 (2003);
%
R.~E.~Browne and J.~A.~Gracey,
Phys.\ Lett.\ B \textbf{597}, 368 (2004);
%
J.~A.~Gracey,
Eur.\ Phys.\ J.\ C \textbf{39}, 61 (2005);
%
E.~Ruiz Arriola, P.~O.~Bowman and
W.~Broniowski,
Phys.\ Rev.\ D \textbf{70}, 097505 (2004);
T.~Suzuki, K.~Ishiguro, Y.~Mori and T.~Sekido,
Phys.\ Rev.\ Lett.\ \textbf{94}, 132001 (2005);
%
M.~N.~Chernodub {\it et al.},
Phys.\ Rev.\ D {\bf 72}, 074505 (2005).
%
E.G.S. Luna, A.A. Natale, Phys. Rev. {\bf D73},074019 (2006);
%
A. C. Aguilar and J. Papavassiliou, Eur. Phys. J {\bf A31}, 742 (2007).
%
\bibitem{SAULI}
V. Sauli, J. Adam, Jr. , P. Bicudo, Phys. Rev. {\bf D75}, 87701 (2007), ArXive: hep-ph/0607196;
P. Bicudo,  Phys. Rev. {\bf D69}, 074003 (2004).

\bibitem{FAKTNE}
D. Atkinson and D. W. Blatt, Nucl. Phys. B{\bf 151},342 (1979);
C. J. Burden, C.D. Roberts and A. G. Williams, Phys. Lett. B{\bf285}, 347 (1992);
G. Krein, C.D. Roberts and A. G. Williams,
Int. J. Mod. Phys. A{\bf 7}, 5607 (1992);
U. Habel, R. Konning, H. G. Reusch, M. Stingl, S. Wigard,
 Z. Phys. A{\bf336},423 (1990);
M. Stingl, Z. Phys. A{\bf 353},423 (1996);
 P. Maris, Phys. Rev. D{\bf 52},6087 (1995);
 V. N. Gribov, Eur. Phys. J. C{\bf10},91 (1999).
%
\bibitem{BSBCW2004}
 C. M. Bender, Sebastian F. Brandt, J-H Chen, Q. Wang,
 Phys. Rev. D{\bf71}; 025014 (2005).
%
\bibitem{BPRT2003}
M. S. Bhagwat, M. A. Pichowsky, C. D. Roberts, P. C. Tandy,
Phys. Rev. C{\bf68}, 015203 (2003);
%
\bibitem{ALDEFIMA2003}
 R. Alkofer, W. Detmold, C. S. Fischer, P. Maris
Comments: 42 pages, 16 figures, revtex; version to be published in Phys Rev D
Journal-ref: Phys.Rev. D70 (2004) 014014
%
\bibitem{BPT2003}
M. Bhagwat, M. A. Pichowsky, P. C. Tandy, Phys. Rev. D{\bf67},054019 (2003).
%
\bibitem{TDM2003}
B. C. Tiburzi, W. Detmold, G. A. Miller, Phys. Rev. D{\bf68},073002  (2003).
%
\bibitem{SUDARSAN}
A.M. Gleeson, R.J. Moore, H. Rechenberg, E.C.G. Sudarshan,
Phys. Rev. {\bf D4} ,2242 (1971);
%
N. Nakanishi, Phys. Rev. {\bf D3},811 (1971).
%
\bibitem{NEWNAKAN}
N. Nakanishi, hep-th/0609206.



\end{thebibliography}
\end{document}